\documentclass[12pt]{article}
\usepackage{epsfig}

\usepackage{graphicx}
\usepackage{epstopdf}
\catcode`\^^?=9 % ignore

\def\half{\mbox{\small{$\frac{1}{2}$}}}

\def\quarter{\mbox{\small{$\frac{1}{4}$}}}
\def\mfrac#1#2{\mbox{\small{$\frac{#1}{#2}$}}}

\def\be{\begin{equation}}
\def\ee{\end{equation}}
\def\ba{\begin{eqnarray}}
\def\ea{\end{eqnarray}}
\def\nn{\nonumber}
\def\ban{\begin{eqnarray*}}
\def\ean{\end{eqnarray*}}
\def\mref#1{Eq.(\ref{eq:#1})}
\def\nref#1{(\ref{eq:#1})}
\def\mlab#1{\label{eq:#1}}

\begin{document}

\begin{center}

{ \Large Nonconservation of Energy and Loss of Determinism}
 \vspace{4mm}
 
 {\large I. Infinitely many colliding balls}
 \vspace{4mm}
  
 {\small David Atkinson, 
 University of Groningen, 9712 GL   Groningen, The Netherlands}

{\small Porter Johnson, Illinois Institute of Technology, Chicago, IL 60616, U.S.A.}

\end{center}

\noindent
{\bf Abstract}

\noindent
An infinite number of elastically colliding balls is considered in a classical, and then in a relativistic setting. Energy and momentum are not necessarily conserved globally, even though each collision does separately conserve them. This result holds in particular when the total mass of all the balls is finite, and even when the spatial extent and temporal duration of the process are also finite. Further, the process is shown to be indeterministic: there is an arbitrary parameter in the general solution that corresponds to the injection of an arbitrary amount of energy (classically), or energy-momentum (relativistically), into the system at the point of accumulation of the locations of the balls. Specific examples are given that illustrate these counter-intuitive results, including one in which all the balls  move with the same velocity after every collision has taken place.

\section{Introduction}

In 5th century B.C. Greece an acrimonious discussion raged. Is the multiplicity of things real or is it a delusion? Are there many things or is there in reality only one thing? Is motion real or merely apparent? Zeno of Elea defended his master, Parmenides, who had asserted that there is truly one indivisible thing and  all motion is illusory. Embarrassingly naive though this debate may sound to modern ears, its ramifications have borne rich fruit during two millennia and a half. Philosophers, mathematicians and physicists have gained much by unravelling the paradoxes of Zeno.  

The paradox of motion usually called the `Achilles' has attracted especial attention.  
A tortoise begins at a point 1, and it crawls along a straight line, passing successively the points $\half$, $\quarter$, $\mfrac{1}{8}$ etc., in some suitable units of distance. Achilles begins at the point  2, and he reaches 1 when the tortoise is at point $\half$,  he reaches $\half$ when the tortoise is at $\quarter$,  and so on. Calling the spatial positions 1,  $\half$, $\quarter$, $\mfrac{1}{8}$, etc., `Zeno points' and the spaces between them `Zeno intervals', we summarize the essence of the paradox: whenever Achilles is at a Zeno point, the tortoise is at the next Zeno point. Since there is no end to the Zeno points, and thus no end to the situation in which Achilles lags behind the tortoise, Achilles cannot draw level with, let alone pass the animal: he will always be one Zeno-interval behind.
What is wrong with this account? Since the 19th century this question has been answered in a way that has satisfied most mathematicians and many  philosophers\cite{Broad}. It runs as follows. The fact that Achilles draws level with the Zeno points one after the other does not imply that he attains only these Zeno points (and presumably covers the Zeno intervals between them); he might reach other points as well. The rules specified by Zeno --- let us call them `Zeno rules' --- are silent about any other points that Achilles might attain. A fortiori they say nothing about whether or not Achilles might reach the particular non-Zeno point that is the limit of all the Zeno points (namely 0). 

The above analysis might seem open to the following objection: `Achilles is  perhaps free to leave the Zeno path of Zeno points and Zeno intervals, but he is only allowed to do so after he has finished the entire journey. And since the Zeno rules dictate that the path consists in an infinite number of Zeno intervals, Achilles will be unable to finish this (super)task. Hence he is not really free, and in this sense the Zeno rules do entail that Achilles is forever confined to the Zeno intervals, simply because there is no end to them.'
This objection calls upon a common intuition, namely that the completion of an infinite number of tasks is impossible because it would take an infinite amount of time. However, this intuition is mistaken, as can be appreciated by considering the set of successive Zeno times, $t_n$, at which Achilles reaches the spatial Zeno points, $s_n$. These successive times form a bounded, monotonic sequence of numbers, if, for example, Achilles runs at a constant speed. Indeed the sequence of times, $t_n$, converges to a limit. Although a vague idea of convergence and limits was entertained in earlier times, it was only with the work of Cauchy, Weierstra{\ss} and Dedekind in the nineteenth century that the problem was finally laid to (mathematical) rest. The bounded, monotonic sequence of times $\{ t_n\}$ satisfies Cauchy's condition
\[
\forall \varepsilon >0,\hspace{8mm}\exists n\in N:\hspace{8mm}\forall p>n\hspace{4mm}\&\hspace{4mm}\forall q>n,\hspace{8mm}|t_p-t_q|<\varepsilon\,.
\]
Roughly, the Zeno times, $t_n$, get indefinitely close together; and the above expression is a definition of a Cauchy sequence. It can be proved that a real finite number, $t$, exists that is the limit of such a monotonic Cauchy sequence, i.e. a number for which the following is true:
\[
\forall \varepsilon >0,\hspace{8mm}\exists n\in N:\hspace{8mm}\forall p>n,\hspace{4mm}|t_p-t|<\varepsilon\,.
\]
This limit may in general be an irrational number, although in our case, if Achilles' speed is a rational number, so is $t$. 
The lead that the tortoise has over Achilles is zero precisely at time $t$. This may be called the 19th century solution of the paradox of Achilles and the tortoise and, aside from a few dissident philosophical voices expressing worries about the feasibility of various supertasks, so the matter rested until 1996, when Jon P\'erez Laraudogoitia introduced a new, but related problem\cite{Laraudogoitia}.

\section{Infinite number of colliding balls}
A system containing a finite number of  balls that undergo a finite number of elastic collisions amongst themselves respects the laws of conservation of energy and momentum. However, consider the following idealized system. An infinite number of identical point masses (balls) are placed at the Zeno points  1, $\frac{1}{2}$, $\frac{1}{4}$, $\frac{1}{8}\ldots $ on a straight line. All the balls are at rest except the first, at 1, which moves with constant speed towards the second, at $\frac{1}{2}$ (see Fig. 1). After elastic collision the first ball comes to rest, passing all its kinetic energy on to the second ball, which soon collides with the third ball, which acquires all the energy, and so on {\em ad infinitum}.  However, after the {\em finite} time that it would have taken the first ball to reach the point 0, had the other balls not been in its way, every ball will have  moved briefly, but then have been brought to rest. After all motion has subsided the energy has disappeared without trace. 

\begin{picture}(0,50)(-10,-30)
\put(96,0){$\leftarrow$}
\put(100,0){\sf o}
\put(50,0){\sf o}
\put(25,0){\sf o}
\put(12.5,0){\sf o}
\put(6.25,0){\sf o}
%\put(3.125,0){\sf o}
\put(3,0.9){\tiny ...}
\put(0,-1.22){*}
\put(-0,-15){\sf Figure 1. Collision of an infinite number of identical balls}
\end{picture}

%Not necessarily so if the number of balls is infinite. An infinite number of elastic collisions can lead to loss of energy.
\noindent

The conclusion is that the energy conservation law has been violated, which is worrying. Should one not be allowed to consider point particles for some reason (despite Newton's frequent use of a massive corpuscle --- {\em Lat. corpusculum} --- in his elegant demonstrations\cite{Newton})? Suppose then that the balls are spheres with geometrically decreasing radii, in such a way that they all fit on a finite line segment (see Fig. 2).

\begin{picture}(0,50)(-10,-30)
\put(89,-1){$\leftarrow$}
\put(100,0){\circle{30}}
\put(50,0){\circle{14}}
\put(25,0){\circle{10}}
\put(12.5,0){\circle{5}}
\put(6.25,0){\circle{3}}
\put(2,-0.27){\tiny ...}
\put(0,-2.5){*}
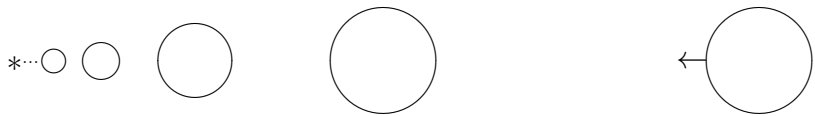
\put(-10,-20){\sf Figure 2. Collision of an infinite number of progressively smaller balls}
\end{picture}

\noindent
If the balls are progressively more dense, in such a way that they all have the same mass, then the analysis goes through unchanged: an infinite number of elastic collisions leads to the loss of  all energy and momentum.

Of course, balls that become denser and denser without limit are grossly unphysical entities. Moreover, an infinite number of identical masses corresponds to an infinite total mass, a theoretical monstrosity. How would it be if the density of the progressively smaller balls were constant, so that the masses decrease geometrically? Now each ball is not brought to rest by collision with its neighbour, so that it retains some kinetic energy after its final collision. Is it possible that the sum of the energies of all the balls, after the infinite sequence of collisions has taken its course, is equal to the initial energy? Indeed that is what happens when the masses decrease in geometrical progression,  at any rate if the rules of Newtonian mechanics are followed. Energy and momentum are conserved and sanity seems to have been restored. Or has it? It turns out that the velocities of the balls are not bounded from above: some of the very tiny balls, indeed all but a finite number of them, acquire speeds in excess of that of light. Evidently the system with geometrically decreasing masses should be treated more properly according to special relativistic mechanics. When this is done, it is found that the tiny balls have speeds close to, but always less than that of light, as expected. However one would also expect the system to obey the law of conservation of energy-momentum, but it does not. While the system does exhibit conservation of energy and momentum classically, this is not so according to relativistic mechanics. These results were obtained earlier by one of us\cite{Atkinson}. 

Apparently relativistic mechanics by itself does not always yield energy-momentum conservation. Of course a physicist can seek an escape from this unpalatable finding by pointing to further unphysical features of the system in question. In particular, the atomic nature of matter requires that the sequence of smaller and smaller balls could not actually be constructed: no gold ball can have a mass less than that of one atom of the lightest isotope of the element. Even if we allow balls to be made of subatomic units,  the ultimately discrete nature of matter will preclude, even in principle, the implementation of a collision scenario involving indefinitely small balls. But this seems like an improper rescue attempt for the conservation law. Are we really prepared to say that the atomic theory of matter is a consequence of the law of conservation of energy-momentum? Surely not. It seems that the bullet must be bitten: when an infinite number of collisions is involved, energy-momentum conservation is indeed not implied by the laws of mechanics. 

These results are interesting from a conceptual point of view. It is known that energy-momentum conservation can fail for systems that are infinite in spatial or temporal extent\cite{Earman}; but what is different in the present case is that the relevant space-time intervals are finite. Nevertheless the `law' of conservation of energy-momentum is not obeyed. There is of course a sort of singularity inherent in the system of balls, namely their infinite number and the fact that  there is a point of accumulation of their locations at the origin of coordinates. How this feature opens the way to the possibility of energy-momentum loss can be seen qualitatively as follows: consider an intermediate time at which the first $n$ balls have experienced one or more collisions, whereas the remainder have not yet been struck. At such a time, energy-momentum is surely conserved, since only a finite number of elastic collisions have taken place. At this intermediate time the energy-momentum can be written as the sum of those of the first $n-1$ balls, and that of the $n$th ball, which will later collide with the $(n+1)$st ball. The crucial question is whether the energy-momentum of this   $n$th ball, {\em after its first but before its second collision}, let us call it inter$(n)$, tends to zero or not as $n$ tends to infinity. If this limit is not zero, then energy-momentum will not be conserved in the complete collision process. This is so because in the limit the sum of the final energy-momenta of all the balls, {\em plus the nonzero limit} of  inter$(n)$, is equal to the initial energy-momentum. In the original example of  identical balls, 
at the intermediate time the   $n$th ball is moving while the first $n-1$ have already been brought to rest, so inter$(n)$ is equal to the initial energy-momentum. If the successive masses become smaller and smaller, on the other hand, the energy-momentum will be partitioned between the first $n-1$ balls and the 
$n$th ball; and whether inter$(n)$, the intermediate energy-momentum of this $n$th ball, has a non-zero limit or not is far from obvious: the answer depends on the rate of decrease of the masses of the balls, and upon whether we do the calculation according to classical or relativistic mechanics. Loosely, one can say that some of the energy may be `lost in an accumulation point'.

A familiar use of infinitude in statistical mechanics is not to engineer energy loss, but rather to simulate irreversibility. A finite, enclosed, elastic system can have no real irreversibility but at best Poincar\'e recurrence on a transcosmic timescale. Formal irreversibility requires the taking of an infinite limit; and similarly the loss of energy occurs only as one lets the number of collisions tend to infinity. 
It has been conventionally assumed that time reversal invariance holds for a system of an infinite number of colliding balls, but can we be certain that this is true? After all, energy is conserved for a finite number of elastic collisions, but not in general for an infinite number. A priori one might entertain the idea that, while the time-reversed version of a finite history of collisions is in accordance with the laws of mechanics, this invariance might not extend to the infinite case.  A system of an infinite number of colliding balls seems at first sight to have a well-defined solution when one ball is initially moving and energy is transmitted along the line of balls, but in general a definite fraction of the energy is lost. 
However, this is not all, since  motion may originate from the accumulation point  and travel towards the right, i.e. back to the first ball. 
To understand intuitively how this can happen, consider first a large but finite number, $N$, of balls. Clearly if we were free to impart some positive momentum and energy to this last, $N$th ball, it would collide with the $(N-1)$st ball, initiating a chain of collisions towards the right. Now imagine increasing $N$, but keeping the imparted energy fixed. In the limit we obtain a chain of collisions towards the right, {\em without any initiating collision, i.e. without any cause}. Whether this limit can be defined is a mathematical question; we shall show below that generally it does make sense. There are solutions of the homogeneous equations that are parametrized by one real number, which may be equated to the energy that spontaneously arises in the system. If this parameter is chosen to be equal to the energy lost in the direct process (if any), then the `forward' history of collisions is precisely repeated, in reverse temporal order, as in a film played backwards. A mechanical system is said to be time-reversal invariant if the time-reversed transform of any solution of the equations of motion is also a solution of those equations. In this sense time-reversal invariance does hold for the infinite system of colliding balls. There exists also an infinite set of alternative solutions, and one needs a boundary condition, namely a specification of the energy lost or gained at the accumulation point, in order to achieve uniqueness.

This paper presents, for the first time, a general solution for the one-dimensional collision problem of the Zeno type. It contains two main sections: in the first classical, and in the second  special relativistic mechanics are used. The structure of these two sections is similar: first the forward, and then the reverse processes are calculated in a general setting. Then special attention is devoted to the situation in which the final velocities of the balls are  all the same. 
The existence of a homogeneous solution provides the mathematical basis for the lack of determinism of the system; it also underpins the time-reversal invariance of the collision process.
In a concluding section the basic reason for the strange results is located in the occurrence of an {\it open} set of mass points; and a further difficulty is mentioned, but its solution is relegated to a future paper.  

%\end{document}

\section{Nonrelativistic collisions}
\subsection{Forward Zeno process}

At each Zeno point, $x_n=2^{-n}$,  $n=0,1,2,\ldots$, there is a ball of mass $m_n$. All the balls are at rest except for the  zeroth one, and that has velocity $\beta_0$ towards the left, causing it to collide with the first Zeno ball, starting an infinite sequence of collisions. Let $u_n$  be the velocity of the $n$th Zeno ball just before its last collision, and $v_n$ its velocity after its last  collision. If $m_{n+1}<m_n$ for all $n$, the zeroth ball has only one collision, while all the other balls undergo two collisions, first one from the right and then one from the left. 

Conservation of momentum and kinetic energy for the collision between the $n$th and the $(n+1)$st balls are expressed by 
\ban
m_nu_n&=&m_nv_n+m_{n+1}u_{n+1}
\\   \half m_nu^2_n&=& \half m_nv^2_n+ \half m_{n+1}u^2_{n+1}\,,
\ean
and with the notation  
\[
\mu_n =\frac{m_{n+1}}{m_n}
\]
these equations lead to 
\be
v_n =u_{n+1} - u_n =u_n - \mu_n \, u_{n+1} \mlab{con}\,,
\ee
and thence to the recurrence relation 
\ba
u_{n+1} &=& \frac{2}{1 + \mu_n} \, u_n \mlab{2nonrel} \\
v_n &=& \frac{1 - \mu_n}{1 + \mu_n} \, u_n  \,.
\nn 
\ea
Since $\mu_n < 1$ it follows that $v_n$ is positive and that $u_{n+1}$ is greater than  $v_n$
(we count velocities to the left as positive for this forward process).

The iterative solution of \nref{2nonrel} is
\ba
u_n &=& \beta_0 \, \prod_{k=0}^{n-1} \frac{2}{1 + \mu_k}\mlab{2nonrela}
\ea
for $n=1,2,\ldots$, where $\beta_0$ is the initial velocity of the zeroth ball. Since 
\ban
m_n &=& m_0 \, \prod_{k=0}^{n-1} \mu_k \,,
\ean
the momentum and energy of the $n$th ball, after its first  collision, but before its second collision, 
can be written in the form
\ban
P_n = m_n u_n &=& m_0\beta_0\prod_{k=0}^{n-1} \frac{2 \, \mu_k}{1 + \mu_k} \\
2T_n = m_n \, u_n^2 &=& m_0  \beta_0^2 \,
\prod_{k=0}^{n-1} \frac{4 \, \mu_k}{(1 + \mu_k)^2}\,.
\ean
Moreover 
\[
T_n=\frac{4\mu_{n-1}}{(1+\mu_{n-1})^2}\, T_{n-1}<T_{n-1} <\ldots <  T_0 \,,
\]
and so
\be
P_n^2 = 2 m_n \, T_n < 2 m_n T_0\mlab{tee}\,.
\ee
If the  total mass $M = \sum_n \, m_n$ 
is finite, it must certainly be the case that $m_n \rightarrow 0$; and then we see from 
\nref{tee} that $P_n \rightarrow 0$ as $n \rightarrow \infty$.  However, $T_n$ does not necessarily
vanish in the limit. 

After the $n$th Zeno ball has suffered its first, but before it has undergone its second collision, the total momentum and kinetic energy are 
\ba
P&=&\sum_{p=0}^{n-1}m_pv_p+P_n=m_0\beta_0
\nn  \\   2T&=&\sum_{p=0}^{n-1}m_pv^2_p+2T_n=m_0\beta^2_0 \mlab{finite}\,.
\ea
In the limit $n\rightarrow\infty$ (which corresponds to the elapse of  only a finite time),  
\ba
P&=&\sum_{p=0}^\infty m_pv_p+\lim_{n\rightarrow\infty}P_n=m_0\beta_0
\nn  \\   2T&=&\sum_{p=0}^\infty m_pv^2_p+2\lim_{n\rightarrow\infty}T_n=m_0\beta^2_0\mlab{infinite}\,.
\ea
Evidently momentum is always conserved when the total mass is finite, for $P_n\rightarrow 0$ as $n\rightarrow\infty$, as we have seen. Kinetic energy is on the other hand conserved only if $T_n\rightarrow 0$ as $n\rightarrow\infty$: that this is not always the case will be illustrated later by means of some examples.  

\subsection{Reverse Zeno process}

We shall now consider the behaviour of this system of balls in the
time-reversed situation in which the velocities of the balls are all
reversed at some time after the completion of the forward Zeno process. 
Since most velocities in the reverse process are in the opposite direction to those for the forward process, it is convenient to switch the sign convention and thus count velocities in the reverse process as being positive if they are to the {\em right,} whereas in the direct process, velocities to the left are counted as positive.

At the instant of reversal the $n$th ball has velocity $w_n$, which is the reverse of its final velocity as calculated during the forward process. One solution of the equations of motion is the precise inverse of the forward process, but it is not the only solution. To see this, suppose that  the $(n+1)$st ball has velocity $u_{n+1}$ at some later time, after which it collides with the $n$th ball, the velocity of which is still $w_n$. As a result of this collision the $(n+1)$st ball acquires a velocity $v_{n+1}$, while the $n$th ball's velocity is changed to $u_n$. Let us summarize the situation for clarity:  the $(n+1)$st ball initially has velocity $w_{n+1}$, it suffers a collision from its left, after which its velocity changes to $u_{n+1}$, then it collides with the ball to its right, after which  its final velocity is $v_{n+1}$. Conservation of momentum and energy at the collision between the $(n+1)$st and the $n$th balls are expressed by   
\ban
m_{n+1}u_{n+1}+m_nw_n&=&m_{n+1}v_{n+1}+m_nu_n
\\   \half m_{n+1}u^2_{n+1} +\half m_nw^2_n&=& \half m_{n+1}v^2_{n+1}+\half m_nu^2_n\,.
\ean
Before a collision can take place, one must have $u_{n+1} > w_n$.
Note that the velocities $u_n$ and $v_n$ are not the same as those in the forward process of the previous section, 
although we use the same symbols.   

The conservation equations yield
\be
u_n=u_{n+1} + v_{n+1}- w_n =\mu_n u_{n+1} + w_n -\mu_n v_{n+1}\mlab{conserv}\,,
\ee
from which we obtain 
\ba
u_n &=& \frac{2 \mu_n u_{n+1} + ( 1 - \mu_n) w_n}{1 + \mu_n} \mlab{2back} \\
v_{n+1} &=& u_n - u_{n+1} + w_n =
\frac{2 w_n- (1 - \mu_n) u_{n+1}}{1 + \mu_n}\,.
\nn
\ea
We shall show that the recursion relation \nref{2back} has a one-dimensional infinity of solutions, of which one is the precise time-inverse of the forward Zeno process. 

The {\it homogeneous} version of \mref{2back} is obtained by setting $w_n=0$, 
\[
\tilde{ u}_n = \frac{2 \mu_n  }{1 + \mu_n}\tilde{u}_{n+1}\,,
\]
and a solution of this is 
\be
\tilde{ u}_0=1\hspace{7mm}{\rm and}\hspace{7mm}
\tilde{ u}_n =  \prod_{k=0}^{n-1}\frac{1+\mu_k}{2\mu_k}\hspace{7mm}{\rm for}\hspace{7mm}n=1,2,3,\ldots\,,
\mlab{homog}
\ee
the general solution being an arbitrary multiple of that. 

Substitute $u_n=f_n\tilde{u}_n$ into \mref{2back} to obtain the recurrence relation 
\[
f_n=f_{n+1}+ \frac{ 1 - \mu_n}{1 + \mu_n}\,\frac{w_n}{\tilde{ u}_n}\,,
\]
with general solution
\be
f_n=\gamma -\sum_{m=0}^{n-1}\frac{ 1 - \mu_m}{1 + \mu_m}\,\frac{w_m}{\tilde{ u}_m}  \mlab{fn}\,,
\ee
where  $\gamma$ is  arbitrary.
The next step is to inject  
\be
w_m =  \beta_0 \,\frac{1 - \mu_m}{1 + \mu_m} \,  \prod_{k=0}^{m-1} \frac{2}{1 + \mu_k}\mlab{inject}
\ee
into  \mref{fn},
which corresponds to the solution \nref{2nonrel}-\nref{2nonrela} of the forward Zeno process. Since $w_m$ for the reverse process is identified with $v_m$ for the forward process, with a change of direction, and because the sign convention for velocities in the reverse process is opposite to that in the forward process, we 
 count $w_m$ as positive for the reverse process when $v_m$ is positive for the forward process.

Accordingly, the general solution of \mref{2back} is
\be
u_n = \left\{ \gamma -\beta_0 \, \sum_{m=0}^{n-1}\left(\frac{ 1 - \mu_m}{1 + \mu_m}\right)^2\,
\prod_{k=0}^{m-1} \frac{4\mu_k}{(1 + \mu_k)^2}\right\}\,\prod_{k=0}^{n-1}\frac{1+\mu_k}{2\mu_k}\,.
\mlab{finalun}
\ee
We recall that $\beta_0$ is the velocity of the zeroth ball at the beginning of the forward Zeno process, and that $\gamma$ is an arbitrary parameter associated with the freedom to add an arbitrary multiple of the homogeneous solution.

\subsection{Example}

As a specific Zeno configuration, consider the mass sequence
\be\mlab{spec}
m_n = \frac{24\, m_0}{(n+1) (n+2) (n+3) (n+4) }\,,
\ee
so that $\mu_n=(n+1)/(n+5)$. The total mass is
$M = \sum_n m_n=\frac{4}{3} \, m_0$. 
For the direct Zeno process we obtain
\ban
u_n &=& \frac{(n+3) (n+4) }{12} \,\beta_0 \\
v_n &=& u_{n+1} - u_n =  \frac{n+4}{6}\, \beta_0\,,
\ean
and no further collisions occur, $v_n$ being positive and monotonically
increasing in $n$.  The energy `lost in the accumulation point' is calculated from 
\[
2 T_n =\frac{(n+3) (n+4) } { 6 (n+1) (n+2)}\,  m_0 \beta_0^2  
\rightarrow \frac{1}{6}\, m_0 \beta_0^2\,,
\]
so one-sixth of the initial energy disappears.

To see that these results are not entirely without physical interest, consider a Zeno system of 101 balls, instead of an infinite number of them. Suppose that $m_0=1$ kg and $u_0=1$ m/s, and that the masses decrease as in \mref{spec}. We find that the last ball, which has a mass of $0.22$ milligrams, carries off 17\% of the initial energy, travelling at 3,200 km/hour!

For the reverse Zeno process we find from Eqs.\nref{homog} and \nref{inject} that 
\ban
\tilde{u}_n&=&\half (n+1)(n+2)
\\   w_n &=& \mfrac{1}{6}\, (n+4)\,\beta_0 \,,
\ean
and therefore 
\[
u_n=\mfrac{1}{12}(6\gamma -5\beta_0)\, (n+1)(n+2)+\mfrac{1}{6}\beta_0\, (2n+5)\,.
\]
From the second of the equations \nref{2back} we obtain for the velocities at the end of the reverse Zeno process 
\[
v_{n+1}= u_n - u_{n+1} + w_n =(\beta_0-\gamma )\, (n+2) \hspace{7mm} {\rm for} \hspace{7mm}n=0, 1,2,3,\ldots\,.
\]
 If $\gamma = \beta_0 $ all balls except the zeroth one are at rest
after this round of collisions.  Moreover $u_0=\beta_0$, and the motion is the exact
time-reversal of the initial collision sequence.  For $\gamma < \beta_0$
the final speeds $v_n$ are positive and increasing in magnitude with
$n$, so that no further collisions can occur.  However, with $\gamma > \beta_0$ additional collisions will take place.
The nature of these collisions, which depend crucially upon spatial locations as 
well as velocities, has not been analyzed. ÊIt is expected to be quite complicated.

\subsection{Constant recoil}

In this section rather special collision sequences are
considered, in which all balls proceed after the second
round of collisions in ``lock step'' with each other at the
same speed $v$,  subsequent collisions being excluded.

The non-relativistic equations reflecting energy and momentum
conservation for the second collision of the $n$th
mass were given in \mref{2nonrel}.  
Under what conditions is $v_n=v$ independently of $n$?
With the notation 
\[
\alpha_n = \frac{1 + \mu_n}{1 - \mu_n}\,,
\]
we deduce from  \mref{2nonrel} that 
\ban
\alpha_{n+1} &=& \alpha_n + 1
\\   u_n  &=& \alpha_n\, v \,.
\ean
The general solution has the form 
\ba
\alpha_n &=& \lambda + n 
\nn   \\ 
u_n &=& (\lambda + n) \, v\,,\mlab{unforward}
\ea
where $\lambda$ is arbitrary. 
The mass ratios are therefore 
\be
\mu_n = \frac{\alpha_n - 1}{\alpha_n +1} =
\frac{\lambda + n - 1}{\lambda + n + 1}\,.
\mlab{constantB}
\ee
From this relation we see that $\lambda >1$ necessarily, since $\mu_0 ={m_1}/{m_0}$ would  otherwise be zero or negative, which would not make sense.

The initial condition may be used to determine $\lambda$ in terms of the original velocity $\beta_0$ of the zeroth ball:
\be
\lambda = \frac{\beta_0}{v}\,,
\mlab{constantD}
\ee
the velocities of the other balls after their first collisions being  
\[
u_n = \beta_0 + n \, v =  \beta_0 \left( 1+ \frac{n}{\lambda}\right) .
\]
We may use \mref{constantB} to determine the masses themselves:
\be
\frac{m_n}{m_0} = \prod_{k=0}^{n - 1} \, \mu_k 
= \prod_{k=0}^{n - 1} \, \frac{\lambda + k - 1}{\lambda + k + 1}
= \frac{\lambda \, (\lambda -1)} {(\lambda + n) \, (\lambda + n - 1)}\,;
\mlab{constantC}
\ee
and so the total mass $M$ of all the balls is 
\be
M = \sum_{n=0}^{\infty} m_n =
\lambda \, (\lambda - 1) \, \sum_{n=0}^{\infty} \, 
\left( \frac{1}{\lambda + n - 1} - \frac{1}{\lambda + n} \right)  m_0
=  \lambda\, m_0\,.
\mlab{constantE}
\ee
The initial momentum and kinetic energy  are given by 
\[
P = m_0 \, \beta_0 
\hspace{25mm} 
2 T = m_0 \, \beta_0^2
\]
and from \nref{constantD} and \nref{constantE} we deduce  their final values to be  
\ban
P_f &=& \sum_{n=0}^{\infty} \, m_n \, v = M \, v
= \lambda  m_0 \,  \frac{\beta_0}{\lambda}
= m_0  \, \beta_0
\\    && \hspace{-20mm} 2T_f = \sum_{n=0}^{\infty} \, m_n \, v^2 =
 M \, v^2 
= \lambda  m_0  \,
\left( \frac{\beta_0}{\lambda} \right)^2 = \frac{m_0 \, \beta_0^2}{\lambda} \,,
\ean
so momentum is conserved but in general energy is not. 

We can try to  gain some insight into the nature of this process by
comparing it to a simple totally inelastic collision, in which a mass
$m_0$ moves with velocity $\beta_0$ toward a mass $M-m_0$ that is initially
at rest.  After the collision the masses stick together and move
in the same direction with speed $v$. Momentum is conserved in this
collision, 
\[
M  \, v =m_0 \, \beta_0 \,,
\]
but energy is not. Indeed, the energy after the collision is 
given by 
\[
2T_f= M\, v^2 =\frac{m^2_0}{M}\beta_0^2=\frac{m_0\beta_0^2}{\lambda} \,,
\]
with $\lambda =M/m_0$, precisely as in  the case of the colliding balls.  
The {\em inelastic} collision of a projectile with a macroscopic, stationary 
target has the same outcome as the infinite number of {\em elastic} collisions,
in the special case that all the masses move with the same final speed after the collisions.
Indeed, the infinite collision sequence can then be considered to be a microscopic
rendition of the single inelastic collision.

\subsection{Constant recoil and time reversal}
Let us now consider the time-reversed version of the above constant recoil scenario in which, at some time
after the collisional process has ended, all velocities are reversed.
The masses $m_n$ are all moving with the same velocity, $w_n =v$. One solution is simply that the backward-moving balls continue in lock step at this speed, and with no further collisions. The general solution of the reverse Zeno process was however calculated in \mref{finalun}, and we simply have to  substitute the particular mass ratio \nref{constantB} into that solution. 
The general solution \nref{finalun} takes on the form 
\ba
u_n &=& \gamma\,\frac{ \lambda +n-1}{\lambda -1}-\beta_0 \frac{\lambda +n-1}{\lambda -1}\nn
 \, \sum_{m=0}^{n-1}\left(\frac{ 1 }{\lambda +m-1}-\frac{ 1 }{\lambda +m}\right)
 \\   &=&\gamma +\frac{n}{\lambda -1} \left( \gamma - \frac{\beta_0}{\lambda }\right)\,.
\mlab{finaluncon}
\ea
In obtaining \mref{finaluncon} we have assumed that the $(n+1)$st 
ball has sufficient velocity to overtake the $n$th ball. Each ball has the same velocity, after the time-reversal but before any further collision has taken place, namely  
\[
w_n=v=\frac{\beta_0}{\lambda}\,,
\]
and from \mref{finaluncon} we deduce that 
\[
u_{n+1}=\frac{\lambda +n}{\lambda -1} \left( \gamma -v\right) +v\,,
\]
so if $ \gamma >v$ then $u_{n+1}>v$ and thus, under this restriction on the otherwise arbitrary $\gamma$, the $(n+1)$st 
ball indeed has  sufficient velocity to overtake the $n$th ball.

From \mref{conserv} we see that 
\be
v_{n+1}=u_n-u_{n+1} + w_n =-\frac{1}{\lambda -1} \left( \gamma - \frac{\beta_0}{\lambda }\right) +\frac{\beta_0}{\lambda }
\; =\; \frac{\beta_0-\gamma}{\lambda -1}\,,
\ee
and thus we conclude that the velocities of the $n$th ball ($n \ge 1$) are
all the same after the final collision.  The zeroth ball ($n=0$), which
does not experience a second collision, has final velocity $u_0=\gamma$.  
%\newpage

The total momentum after all collisions have taken place is 
\ban
m_0 \, u_0 + \sum_{n=1}^{\infty} m_n \, v_n &=&
m_0 \, \gamma + (M-m_0)\, v_n  \; =\; 
   m_0 \, \beta_0 \,,
\ean
and this is equal to the original momentum at the very beginning of the forward collisional process. Momentum is conserved in the backward, as in the forward scenario.

The kinetic energy of the system after all collisions have taken place is not in general conserved. 
Twice the final kinetic energy is  
\ban
 m_0 \, u_0^2 + \sum_{n=1}^{\infty} \, m_n \, v_n^2 &=&
 m_0 \, \gamma^2 +  (M-m_0)\, v^2_n  \; =\; 
 \frac{m_0\,\beta_0^2}{\lambda} +
\frac{m_0 \, \lambda}{\lambda-1} \left( \gamma - v \right)^2\,.
\ean
 Twice the {\it minimum} kinetic energy is $m_0\,\beta_0^2/\lambda$,
which is the value at the end of the  forward Zeno process, and this occurs if $\gamma$ is chosen to be $v$, the lower limit of the permitted  values 
for this variable. By making $\gamma$ larger, the kinetic energy becomes larger, and it can be made even larger than the original kinetic energy at the beginning of the forward process. 

The additional energy in the system is  equal to the energy added at the instant of time reversal.  The time-reversed
system is indeterminate, and the arbitrary parameter
$\gamma$, which is a measure of the energy injected at the origin of coordinates (the accumulation
point of the Zeno points) apparently arises in the dynamics.  Moreover, the reverse Zeno process can start at any moment in time. For that matter, energy could equally well have been injected during the forward Zeno process. This can be envisaged as follows: consider a time after the initiation of the forward process, but before the wave of collisions has run its course to the origin.  At this time, imagine that a spontaneous wave of motion starts at the origin. Since infinitely many balls are still at rest, the solution of interest is a multiple of the homogeneous solution   
\nref{homog}. In fact, the velocities can be read off from \nref{finaluncon} by putting $\beta_0$ equal to 0, since this mimics the situation in which the reverse Zeno process occurs, after a forward process in which nothing happens, i.e. in which all the balls remain at rest. If we also replace $n$ by $n+1$, we find 
\be
u_{n+1}=-\gamma\,\frac{\lambda +n }{\lambda -1 }
\ee
as the velocity of the $(n+1)\,$st ball, after the $(n+2)\,$nd ball has collided with it, but before it collides with the $n$th ball. A minus sign has been added, because for the present calculation we wish to specify velocities toward the left as being positive.  
Now suppose that, just before the latter collision takes place, the wave of collisions starting from the zeroth ball, that originally had velocity 
 $\beta_0$, has reached the  $n$th ball. The velocity, $u_n$, of this ball is then as in \mref{unforward}. After its collision with the $(n+1)\,$st ball, the velocity of the  $n$th ball  becomes 
 \[
 v_n=\frac{(1-\mu_n )u_n+2\mu_n u_{n+1}}{1+\mu_n}\,,
 \]
where $u_{n+1}$ and $u_n$ have just been specified, and the mass ratio, $\mu_n$, was given in \mref{constantB}. We find 
\[
v_n=v-\left( 1+\frac{n}{\lambda -1}\right) \gamma\,,
\]
where we recall from \mref{constantD} that $v = {\beta_0}/{ \lambda}$. From this it follows that $v_n$ can be anything from $v$ to minus infinity, and indeed it is clear without detailed calculation why this is so. For the   $n$th ball acquires a final velocity $v$ on condition that the $(n+1)\,$st ball is at rest when the 
 $n$th ball hits it; but the $(n+1)\,$st ball may have any velocity to the right, as a result of spontaneous generation from the accumulation point, and so in fact any velocity between $v$ and minus infinity is consistent with the dynamical equations.

\section{Relativistic collisions}
\subsection{Forward Zeno process}
What difference does special relativity make to the conclusions of the previous section? 
Conservation of relativistic energy and momentum at the collision between the $n$th and the $(n+1)$st balls is expressed by 
\ban
\gamma (u_n)\, m_n+m_{n+1}&=&\gamma (v_n)\, m_n+\gamma (u_{n+1})\, m_{n+1}
\\   \gamma (u_n)\, m_nu_n&=&\gamma (v_n)\, m_nv_n+\gamma (u_{n+1})\, m_{n+1}u_{n+1}\,,
\ean
where $\gamma (v)$ is the Lorentz factor $(1-v^2)^{-\frac{1}{2}}$, units having been chosen so that the speed of light in vacuo is unity.
On adding and subtracting these two equations, one deduces  
\ban
\epsilon^{-1}(v_n) - \epsilon^{-1}(u_n) &=&
\mu_n \left[ 1 - \epsilon^{-1}(u_{n+1}) \right] \\
\epsilon(v_n) - \epsilon(u_n) &=& \mu_n \left[ 1 - \epsilon(u_{n+1})
\, \right]\,,
\ean
where 
\be
\epsilon(v) = (1-v)\gamma (v)=\sqrt{\frac{1-v}{1+v}} 
\mlab{relco11A}
\ee
and $\mu_n = m_{n+1}/m_n$ as before. 
Equivalently, we have
\be
\epsilon(v_n)=\frac{\epsilon(u_{n+1})}{\epsilon(u_n)} =
\frac{\mu_n + \epsilon(u_n)}{1 + \mu_n \,\epsilon(u_n)}\,,
\mlab{relcollB}
\ee
which is the relativistic generalization of \mref{con}. 
We once more limit consideration to the case in which the total mass $M$ of the
particles is finite, so that $m_n \rightarrow 0$ as
$n \rightarrow \infty$; and we again require the masses $m_n$ to be
monotonically decreasing in $n$, i.e. $\mu_n < 1$. 
We assume $u_0 >0$, so that $\epsilon(u_0) < 1$,   and
\be
\frac{\epsilon(u_{n+1})}{\epsilon(u_n)} = 1 -
\frac{(1 - \mu_n) \, [1 - \epsilon(u_n)]}{1 + \mu_n \,\epsilon(u_n)}
< 1 \mlab{newe1}\,,
\ee
from which it follows by induction that $\epsilon(u_n)$ as a monotonically decreasing function of $n$.  
The velocities $u_n$ increase with $n$, and all of the recoil velocities
$v_n$ are positive.

Since $\epsilon(u_n)$ constitutes a monotonically decreasing set of numbers in the interval $[0,1]$, it must have a limit as $n\rightarrow\infty$. If there is a number $\Delta$ such that $ \mu_n\le\Delta <1$ for all $n$, we see from  \mref{newe1} that 
\[
{\epsilon(u_{n+1})} \le \Gamma \, {\epsilon(u_n)}
\]
where, with the notation $\epsilon_0 \equiv \epsilon(u_0)$, 
\[
\Gamma = 1 -
\frac{(1 - \Delta) \, (1 - \epsilon_0)}{1 + \Delta \,\epsilon_0}
< 1 \mlab{newe2}\,,
\]
since $\epsilon_0 <1$. 
It then follows that  $\epsilon(u_n)\le \Gamma^n \epsilon_0$, i.e $\epsilon(u_n)$ tends to zero as $n\rightarrow\infty$. However, even when $\mu_n$ is only bounded nonuniformly by 1 it is often the case that the limit is still  zero. It has been shown\cite{Atkinson} that, if 
\[
\mu_n\le 1-\frac{a}{n+1}
\]
with $a>0$, then $\epsilon(u_n)\rightarrow 0$. 
Since for $a>1$ Raabe's condition is met  for the convergence of the mass series $M = \sum_n m_n$, it follows that 
 $\epsilon(u_n)$ has zero limit when $M$ is finite, but this is true also in more general contexts.  

The relativistic analogue of \mref{finite} is 
\ba
P&=&\sum_{p=0}^{n-1}m_pv_p\gamma (v_p)+P_n
\nn  \\   E&=& \sum_{p=0}^{n-1}m_p\gamma (v_p)+E_n\mlab{relfinite}\,,
\ea
where 
\[
P_n=m_nu_n\gamma (u_n)\hspace{15mm}E_n=m_n\gamma (u_n)\,.
\]
Momentum and energy are lost if $P_n$ and $E_n$ fail to vanish in the limit  $n\rightarrow\infty$. Limiting our interest to the cases in which 
 $\epsilon(u_n)$ tends to zero in this limit, so that $u_n$ tends to unity (light speed), we see that violation of energy-momentum conservation takes place if 
 \be
 m_n\gamma (u_n)=\half m_n [ \epsilon(u_n)+\epsilon^{-1}(u_n)] \rightarrow \kappa \neq 0  \mlab{kappa}\,.
 \ee
This often happens, except for cases in which the masses $m_n$ decrease more rapidly than exponentially in 
$n$, as discussed in Ref.  4.  Energy-momentum is lost even when $m_n$ decreases as a negative exponential of $n$, in contradistinction to the nonrelativistic situation. The special case of
constant recoil velocities $v_n$ --- for which energy and momentum
are always lost --- will be taken up below.

\subsection{Reverse Zeno process}

Consider the behaviour of this system of balls in the
time-reversed situation. 
Conservation of energy and momentum give 
\ban
\mu_n \, \gamma(u_{n+1}) + \gamma(w_n) &=& \mu_n \, \gamma(v_{n+1})
+ \gamma(u_n) \\
\mu_n \, \gamma(u_{n+1}) \, u_{n+1} + \gamma(w_n) \, w_n &=&
\mu_n \, \gamma(v_{n+1})\, v_{n+1} + \gamma(u_n) \; u_n\,,
\ean
and these equations lead to 
\ban
\epsilon^{-1}(u_n) - \epsilon^{-1}(w_n) &=&
\mu_n \left[ \epsilon^{-1}(u_{n+1}) - \epsilon^{-1}(v_{n+1}) \right] \\
\epsilon(u_n) - \epsilon(w_n) &=& \mu_n \left[ \epsilon(u_{n+1} - \epsilon(v_{n+1})
\, \right]\,.
\mlab{3condition}
\ean
We thus obtain
\be
\frac{\epsilon(u_n)}{\epsilon(u_{n+1})} =
\frac{\epsilon(v_{n+1})}{\epsilon(w_n)}   =
\frac {\epsilon(w_n) + \mu_n \, \epsilon(u_{n+1})}
{\epsilon(u_{n+1}) + \mu_n \, \epsilon(w_n)}\,.
\mlab{relcollD}
\ee
The factors $\epsilon(w_n)$, which depend upon the velocity of
the $n$th ball before any collisions have taken place, are supposed specified, as in the nonrelativistic reverse process.  We can
solve \mref{relcollD} by backward iteration, starting
at an asymptotically large value of $n$ with
$\epsilon(u_{n+1}) =  m_{n+1}/(2\sigma )$, where the arbitrary parameter
$\sigma$ has been introduced.

We shall first focus on the special choice $\sigma =\kappa $, where $\kappa$ was
defined in \mref{kappa} as the asymptotic ratio of
 $m_n/(2\epsilon(u_n))$ in the forward iteration.  Under this circumstance,
we obtain $\epsilon(v_{n+1})= 1$ for all $n$ (or $v_{n+1}=0$).
Furthermore, $\epsilon(u_0)$ is equal to $\epsilon_0$, the original value it had just before the time-forward
collisions began.  This case clearly corresponds to the precise time reversal of the time-forward evolution.

For $\sigma > \kappa $, the energy added at the accumulation point is greater
than that lost in the time-forward case, while for $  \sigma < \kappa $
the energy added is less than that initially lost.
For example, with the  mass sequence 
$m_n = 2^{-n}$ and initial factor $\epsilon_0 = 0.4$, we obtain in the forward iteration 
$\kappa = 0.616082$ and
\[
\lim_{n \rightarrow \infty} \epsilon(v_n) = 0.5\,.
\]
Here is a table of values of $\epsilon(v_1)$, $\epsilon(u_1)$, and
$\epsilon(u_0)$ for various values of the parameter $\sigma$.

\[
\begin{array}{|l||l|l|l|}
\hline
\hspace{2.5mm}\sigma  \hspace{2.5mm} & \hspace{2.5mm} \epsilon(v_1) \hspace{2.5mm}&  \hspace{2.5mm}\epsilon(u_1) \hspace{2.5mm}&  \hspace{2.5mm}\epsilon(u_0) \hspace{2.5mm} \\
\hline 
 \hspace{2.5mm}0.01  \hspace{2.5mm} & \hspace{2.5mm} 0.762 \hspace{2.5mm} &  \hspace{2.5mm}0.578 \hspace{5mm} & \hspace{2.5mm} 0.630 \\
 \hspace{2.5mm}0.1 \hspace{2.5mm} &  \hspace{2.5mm} 0.831 \hspace{5mm} & \hspace{2.5mm} 0.481 \hspace{5mm} &  \hspace{2.5mm}0.557 \\
 \hspace{2.5mm}0.616082 \hspace{2.5mm} &  \hspace{2.5mm} 1.0 \hspace{5mm} &  \hspace{2.5mm}0.3 \hspace{5mm} &  \hspace{2.5mm}0.4 \\
 \hspace{2.5mm}1.0 \hspace{2.5mm} &  \hspace{2.5mm} 1.067 \hspace{5mm} &  \hspace{2.5mm}0.240 \hspace{5mm} & \hspace{2.5mm} 0.340 \\
 \hspace{2.5mm}100.0 \hspace{2.5mm} &  \hspace{2.5mm} 1.485 \hspace{5mm} &  \hspace{2.5mm}0.005\hspace{5mm} &  \hspace{2.5mm}0.010 \\
  \hline
\end{array}
\]

\subsection{Constant recoil}
Let us return to the case in  which all balls recoil with the same speed $v$, but now in the relativistic context.  
We have simply to set $\epsilon(v_n)=\eta$, independently of $n$, in \mref{relcollB}, to obtain 
\ban
\epsilon(u_{n+1}) &=& \eta \, \epsilon(u_n) \\
\eta &=& \frac{\mu_n + \epsilon(u_n)}{1 + \mu_n \, \epsilon(u_n)} \,.
\ean
The solution to the first equation may be expressed in terms of
$\epsilon_0 = \epsilon(u_0)$ as
\[
\epsilon(u_n) = \eta^n \epsilon_0\,.
\]
Inserting this relation into the second equation, and  solving for the mass ratios $\mu_n$, we obtain 
\be
\mu_n = \eta\, \frac{  1 - \eta^{n-1} \epsilon_0}
{1 - \eta^{n+1} \epsilon_0 }
\mlab{constantI}
\ee
The masses themselves can then be determined:
\ba
\frac{m_n}{m_0} &=& \prod_{k = 0}^{n - 1} \, \mu_k 
= \eta^n \, \prod_{k = 0}^{n - 1}
\frac{1 - \eta^{k-1} \epsilon_0}
{1 - \eta^{k+1} \epsilon_0 }
\; =\;  \eta^n
\frac{(1-\epsilon_0) \, (\eta - \epsilon_0) }
{(1 - \eta^n\, \epsilon_0) \, (\eta - \eta^n\, \epsilon_0 )}\,.
\mlab{constantJ}
\ea
As in the non-relativistic case, we can calculate the total mass
of all the balls,
\ba
\frac{M}{m_0} &=& \sum_{n=0}^{\infty} \, \frac{m_n}{m_0} \; =\;
(1 - \epsilon_0) \, (\eta - \epsilon_0) \, \sum_{n=0}^{\infty}
\frac{ \eta^n }{(1 - \eta^n \epsilon_0) \, (\eta - \eta^n \epsilon_0 )}
\; =\; \frac{1 - \epsilon_0}{1 - \eta}\,.\nn
\\   \mlab{constantK}
\ea
We use the relations
\[
2 \gamma(v) = \eta^{-1} + \eta
\hspace{10mm} \mbox{and} \hspace{10mm}
2 v \gamma(v) =  \eta^{-1} - \eta
\]
to determine the energy and momentum lost during the
collision, namely 
\ba
\Delta E &=&  m_0 \, \gamma(u_0)  +(M -m_0)-  M  \, \gamma(v) \nonumber \\
  &=& \frac{m_0}{2 \, \epsilon_0 \, \eta} \, (\eta - \epsilon_0) \,
(1 - \epsilon_0) \nonumber \\
    \Delta P &=& 
m_0 \, \gamma(u_0) \, u_0 -  M  \, \gamma(v) \, v 
\; =\; \Delta E\,.
\mlab{constantL}
\ea
Although the equality $\Delta E=\Delta P$ may seem fortuitous in the above calculation, it is not really so, for it can be readily checked that 
\[
\Delta E =\lim_{n \rightarrow \infty} m_n \, \gamma(u_n)=\lim_{n \rightarrow \infty} m_n \, \gamma(u_n)u_n=\Delta P\,.
\]
The loss of energy and the loss of momentum are equal because $u_n$, the limit as $n\rightarrow\infty$ of the velocity of the $n$th ball, after its first but before its second collision, is equal to unity (light speed).

As in the non-relativistic case, we seek insight by considering a
a simple collision of two elementary masses $m_0$ and $M-m_0$.  The mass
$m_0$, initially moving with velocity $u_0$, strikes the mass $M-m_0$,
which is initially at rest.  After the collision the two masses move together 
with the same velocity $v$, and we imagine the balance of the energy-momentum to be carried away  by a collinear electromagnetic
wave with energy $E_{em} = P_{em}$. 

The following relations express energy and momentum conservation in
this relativistic collision:
\ban
m_0 \, \gamma(u_0) + M -m_0 &=&  M \, \gamma(v)+ E_{em} \\
m_0 \, \gamma(u_0) \, u_0 &=&  M \, \gamma(v) \, v + E_{em}\,. 
\ean
We subtract and add these equations to obtain
\ba
m_0 \, \epsilon_0 + M-m_0 &=& M \, \eta \nonumber \\
m_0 \, \epsilon^{-1}_0 + M-m_0 &=& M \,
\eta^{-1}+ 2 \, E_{em}\,,
\mlab{constantM}
\ea
where we have
adopted the notation $\epsilon_0 = \epsilon(u_0)$ and $\eta =
\epsilon(v)$, as in the previous calculation.
The first relation may be written as
\[
M = m_0 \, \frac{1 - \epsilon_0}{1 - \eta}\,,
\]
in agreement with \mref{constantK}.  Furthermore, the second relation yields
\ban
E_{em} &=& 
\half m_0\left( \epsilon^{-1}_0 - \epsilon_0 \right) 
- \half M\left( \eta^{-1} - \eta \right)
 \\
&=& \frac{m_0}{2\, \epsilon_0 \,\eta } \, (\eta - \epsilon_0) 
\, (1 - \epsilon_0) \,.
\ean
The electromagnetic momentum and energy are identical to that lost in
the sequential collision just considered.

As in the non-relativistic case considered in Sect. 3.4,
the energy lost in the collision of the balls of
specified masses is greatest when the balls
move in lock step after the collision.  
These collisions, in which the balls move away with the same speed
after the collisions are completed, have the property that the lost 
mechanical energy-momentum is precisely equal to the energy-momentum that is 
degraded into electromagnetic form.

\subsection{Constant recoil and time reversal}
We once more reverse the velocities of all masses at some time after the
collision sequence has ended. We obtain from \mref{relcollD} 
\be
\epsilon(u_n) = \frac{\eta + \mu_n \, \epsilon(u_{n+1})}
{\epsilon(u_{n+1}) + \mu_n \, \eta} \, \epsilon(u_{n+1})
\mlab{constantN}
\ee
and
\be
\epsilon(v_{n+1}) = \eta \,\frac{\epsilon(u_n)}{\epsilon(u_{n+1})}\,,
\mlab{constantO}
\ee
where, as before, $\eta = \epsilon(v)$.
At large $n$, if follows from \mref{constantI} that
$\mu_n \sim \eta$ as $n \rightarrow \infty$.  We seek solutions
such that that $\epsilon(u_n) \rightarrow 0$ in that limit, so from \mref{constantN} 
we read off the asymptotic recursion formula 
\[
\epsilon(u_{n+1}) = \eta \,\epsilon(u_n)\,.
\]
Thus, for large $n$, 
\[
\epsilon(u_n) \sim \omega \,\eta^n\,,
\]
where the positive parameter $\omega$ may be freely chosen.  We may
determine  $\epsilon(u_n)$ for all $n$ by inserting this asymptotic form
for $\epsilon(u_{N+1})$ at a large value of $N$,  and generating $\epsilon(u_n)$ by
backward recursion of the formula \nref{constantN},
obtaining $\epsilon(v_{n+1})$ at each stage of the iteration by
using \nref{constantO}.  The energy added at the accumulation point
of the masses is determined to be 
\be
\lim_{n \rightarrow \infty} \frac{m_n}{2 \epsilon(u_n)}
= \lim_{n \rightarrow \infty}
\frac{m_0 \, \eta^n \, (1 - \epsilon_0) \, (\eta - \epsilon_0)}
{2 \omega \, \eta \, \eta^n}
= \frac{m_0 \, (1 - \epsilon_0) \, (\eta - \epsilon_0)}
{2 \omega \, \eta}\,.
\mlab{constantP}
\ee
The functions $\epsilon(v_n)$  and $\epsilon(u_0)$ may be determined numerically.
The conditions $u_n > v$, or equivalently $\epsilon(u_n) < \epsilon(v)$, for all $n$, must also be met in order 
 that the initial collisions 
indeed take place.  
For the special choice $\omega = \epsilon_0$, the quantity
\be
\epsilon(u_{n+1}) = \epsilon_0 \, \eta^{n+1}
\mlab{constantQ}
\ee
satisfies the recursion formula \nref{constantN} for all
$n$, and results in $\epsilon(v_n) = \eta$.  

Here is a table of values of $\epsilon(u_0)$ and $\epsilon(v_1)$
for various values of the parameter $\omega$, when $\epsilon_0 = 0.4$
and $\eta = 0.6$, for which $M = m_0/2$.
\[
\begin{array}{|l||l|l|l|}
\hline 
\hspace{2.5mm}\omega\hspace{2.5mm} & \hspace{2.5mm}\epsilon(v_1)\hspace{2.5mm} & \hspace{2.5mm}\epsilon(u_1)\hspace{2.5mm} &\hspace{2.5mm} \epsilon(u_0)\hspace{2.5mm} \\ \hline 
\hspace{2.5mm}0.01\hspace{2.5mm} &\hspace{2.5mm} 2.241\hspace{2.5mm} &\hspace{2.5mm} 0.013\hspace{2.5mm} &\hspace{2.5mm} 0.045\hspace{2.5mm} \\
\hspace{2.5mm}0.1\hspace{2.5mm} & \hspace{2.5mm}1.456\hspace{2.5mm} &\hspace{2.5mm} 0.100\hspace{2.5mm} &\hspace{2.5mm} 0.243\hspace{2.5mm} \\
\hspace{2.5mm}0.4\hspace{2.5mm} &\hspace{2.5mm} 1.000\hspace{2.5mm} &\hspace{2.5mm} 0.240\hspace{2.5mm} &\hspace{2.5mm} 0.400\hspace{2.5mm} \\
\hspace{2.5mm}1.0\hspace{2.5mm} & \hspace{2.5mm}0.826\hspace{2.5mm} &\hspace{2.5mm} 0.344\hspace{2.5mm} &\hspace{2.5mm} 0.473\hspace{2.5mm} \\
\hspace{2.5mm}100.\hspace{2.5mm} &\hspace{2.5mm} 0.615\hspace{2.5mm} &\hspace{2.5mm} 0.576\hspace{2.5mm} &\hspace{2.5mm} 0.589\hspace{2.5mm} \\
\hline
\end{array}
\]
For $\omega = \epsilon_0 = 0.4$, which was discussed
previously, the factors $\epsilon(v_{n+1})$ are equal to unity for all $n$,
and thus the recoil speeds, $v_{n+1} $, are zero.  Also, the factors
$\epsilon(u_{n+1})$ are in agreement with \mref{constantQ}, and
speed of the first mass is the same as its incident velocity
that initiated the collision process in the forward Zeno process.
The energy added (see \mref{constantP}) is identical
to that lost in the forward case (see \mref{constantL},
and thus the motion is the precise temporal inverse of the forward Zeno process.

For $\omega < \epsilon_0$, the quantities $\epsilon(v_{n+1})$ are
greater than unity and monotonically decreasing in $n$, so that the
final speeds are negative and monotonically decreasing in magnitude.
Consequently there are no collisions after the second round. 
For $\omega > \epsilon_0$, on the other hand, the quantities
$\epsilon(v_{n+1})$ are less than $1$, positive, and monotonically
decreasing in $n$, so that the recoil speeds are monotonically
increasing, and subsequent collisions must occur.  As in the non-relativistic case, we have not further analyzed this complicated situation.

\section{Discussion}

The loss of energy --- or energy-momentum in the relativistic case ---  may upset the physicist, and the lack of determinacy is even more bizarre. For indeed this indeterminism is very radical.  Since a spontaneous wave of motion of arbitrary energy may originate from the origin at any time whatsoever, it is strictly undetermined what the velocity of any ball will be from one moment to the next. A ready objection to such a muddled situation is that  the initial condition  has not been fully specified. For a finite number of balls, it is sufficient to specify their positions and velocities at any one time in order to be able to determine their configuration at any later (or earlier) time. For an infinite number of balls, we have shown in this paper that such a specification is insufficient. One must also specify `the energy at any accumulation point' of the positions of the balls. However this is very awkward, for it is not enough to limit the specification to any one time: it would have to be done for all time. A simple solution would be simply to forbid all injection of energy --- perhaps on the grounds that such an injection would be tantamount to intervention from outside the system. However, such a proscription comes at a price. It would mean that the equally massive balls, after they have all come to rest, remain forever at rest. For the case of constant recoil, whether treated classically or relativistically, it would mean that the balls remain forever in lock step and never collide again. This implies however that an arrow of time has been imported by fiat into the specification of the laws of mechanics. While it is notoriously difficult to obtain an arrow of time, this seems too cheap a way to achieve one. 

The physicist would be inclined to reject both the nonconservation of energy-momentum and indeterminism, on the grounds that the `physical' way to treat an infinite system is by way of the finite system, in the limit that the number of subsystems approaches infinity. Thus he might simply reject the analysis of the infinite set of identical balls, since there is no continuity in the infinite limit (all the energy-momentum comes to reside in the last ball, for any finite number of balls, but with an infinite number of them there is no `last' ball).  
In the case that the masses decrease, however, there is a ready way to achieve continuity in the limit. As we have mentioned above, the missing energy-momentum is associated with zero rest-mass. Thus we have only to posit that the missing energy-momentum is carried off as light, in order to restore continuity in the limit and conservation of energy-momentum. The time-reversed motion involves the absorption of light of just the right energy-momentum to ensure that the reverse process exactly mirrors the forward process. The specter of indeterminism is exorcised by requiring that the specification of an initial condition include the  `energy-momentum at the accumulation point'. It remains true of course that the homogeneous version of the equations of motion has a nontrivial solution, and that mathematically any multiple of a homogeneous solution can be added to one inhomogeneous solution to yield another. But it is a familiar situation in mathematical physics that some mathematical solutions of an equation are patently `unphysical'. 

The philosopher is typically unimpressed by ad hoc appeals to what her physicist colleague deems to be physical. For her the law of conservation of energy is not entailed by the laws of mechanics, nor is determinism,  since an infinite number of Zeno balls provides a counter-example. One either accepts this result at face value, or one seeks to exclude the unwelcome conclusion by further circumscribing what counts as an acceptable, albeit still ideal system.  

The above difficulties are all associated with the occurrence of an open set of mass-points. Grave though the difficulties are that have just been  mentioned, there is potentially a  
more serious problem  that has not yet been considered. What would happen if the zeroth ball were to approach the other balls, not from the right of them all, but from the left? That is, what would happen if the zeroth ball starts at some negative position on the $x$-axis and travels towards the right, colliding with the accumulation point of balls at the origin? This question was posed by Alper and Bridger\cite{alper1}, and their answer was that no such collision is logically possible, and that the ball would simply have to disappear!  

In a sequel to this paper, ``Nonconservation of Energy and Loss of Determinism. II. Colliding with an open set'', we intend to analyze this situation in mathematical detail. We will discuss certain less drastic solutions than the one proposed by Alper and Bridger that have been considered in the literature, and we will present our own resolution of the looming paradox.

\end{document}